\documentclass[prd,showpacs,preprintnumbers,nofootinbib,aps]{revtex4}
\usepackage{subeqnarray}
\usepackage{epsfig,amsmath,amssymb}
\usepackage{mathrsfs}
\usepackage[usenames,dvipsnames]{color}
\usepackage[pagebackref=false, colorlinks=false]{hyperref}
\definecolor{redish}{rgb}{0.7,0.2,0.0}  
\definecolor{bluish}{rgb}{0.2,0.5,0.8}
\hypersetup{linkcolor=redish,          
                  citecolor=blue,        
                  filecolor=magenta,      
                  urlcolor=bluish}          
\DeclareFontFamily{U}{rsfs}{}         
\DeclareFontShape{U}{rsfs}{m}{n}{<5> rsfs5 <6><7> rsfs7          %
  <8><9><10><10.95><12><14.4><17.28><20.74><24.88> rsfs10}{}     %
\DeclareMathAlphabet{\mathfs}{U}{rsfs}{m}{n}                     %

\newcommand{\ba}{\nopagebreak[3]\begin{eqnarray}}
\newcommand{\ea}{\end{eqnarray}}
\newcommand{\bii}{\begin{itemize}}
\newcommand{\eii}{\end{itemize}}

\begin{document}

\title{Thermal Fluctuations And Correlations Among Hairs Of Stable Quantum ADS Kerr-Newman Black Hole}
\author{Aloke Kumar Sinha}
\email{akshooghly@gmail.com}
\affiliation{ Ramakrishna Mission Vivekananda University, Belur Math 711202,India\\ Haldia Government College,West Bengal, India  }
       
\begin{abstract}

 We have already derived the Criteria for thermal stability of charged rotating quantum black holes, for horizon areas that are large relative to the Planck area. The derivation is done by using results of loop quantum gravity and equilibrium statistical mechanics of  the Grand Canonical ensemble. It is also shown that in four dimensional spacetime, quantum ADS Kerr-Newman Black hole is thermally stable within certain range of its' parameters. In this paper, the expectation values of fluctuations and correlations among  horizon area, charge and angular momentum of stable quantum ADS black hole are calculated within the range of stability. Interestingly, it is found that leading order fluctuations of charge and angular momentum, in large horizon area limit, are independent of the values of charge and angular momentum at equilibrium.

\end{abstract}
\maketitle

\section{Introduction}

 Semiclassical analysis shows that nonextremal, asymptotically flat black holes are thermally unstable  due to decay under Hawking radiation, with negative specific heat  \cite{dav77}. This motivated the study of thermal stability of black holes, from a perspective that relies on a definite proposal for {\it quantum} spacetime (like Loop Quantum Gravity, \cite{rov,thie}) . A consistent understanding  of {\it quantum} black hole entropy has been obtained through Loop Quantum Gravity \cite{abck98,abk00}, where not only has the Bekenstein-Hawking area law been retrieved for macroscopic (astrophysical) black holes, but a whole slew of corrections to it, due to quantum spacetime fluctuations have been derived as well \cite{km98}-\cite{bkm10}, with the leading correction being logarithmic in area with the coefficient $-3/2$.\\
 
 Classically a black hole, in general relativity, is characterized by its' mass ($M$), charge ($Q$) and angular momentum ($J$).  Intuitively, therefore, we expect that thermal behaviour of black holes will depend on all of these parameters. The simplest case of vanishing charge and angular momentum has been investigated longer than a decade ago \cite{cm04} - \cite{cm05-2} and that has been  generalized, via the idea of {\it thermal} holography \cite{pm07}, \cite{pm09}, and the saddle point approximation to evaluate the canonical partition function corresponding to the horizon, retaining Gaussian thermal fluctuations.  This body of work has been generalized recently \cite{aspm 16} for charged rotating black holes. There it is shown that  anti-de Sitter(ADS) Kerr-Newman black hole (for a certain range of its' parameters) is thermally stable . In fact the conditions for thermal stability of a macroscopic quantum black hole with arbitrary number of hairs in arbitrary spacetime dimension  has already been derived too \cite{aks} . \\

In this paper, using previous knowledge {\cite{aspm 16}, thermal fluctuations and correlations among all the hairs i.e. charge, horizon area and angular momentum are calculated. These are calculated in the limit of large horizon area. \\  

The paper is organized as follows: In section 2, the idea of thermal holography, alongwith the concept of (holographic) mass associated with horizon of a black hole is briefly reviewed along with detail discussion of quantum black hole algebra and quantum geometry. This section also contains a short revision of grand canonical partition function of charged rotating black hole (ADS Kerr-Newman Black Hole) and condition for its' thermal stability. In the next section, detailed calculation of thermal fluctuations are done for ADS Kerr-Newman black hole. Last section contains a brief summary and outlook.

\vspace{.3 in}
\section{Thermal holography}

 In this section, we briefly review some part of our earlier work, essencial for this paper and hence some overlapping with that \cite{aspm 16} is inevitable. 

\subsection{Mass Associated With horizon}

Black holes at equilibrium are represented by isolated horizons, which are internal boundaries of spacetime. Hamiltonian evolution of this spacetime gives the first law associated with isolated horizon($b$) and is given as,
\begin{eqnarray}
\delta E^{t}_{h}=\frac{\kappa^{t}}{8\pi}\delta A_{h}+\Phi^{t}\delta Q_{h}+\Omega^{t}\delta J_{h}
\end{eqnarray}
where, $E^{t}_{h}$ is the energy function associated with  the horizon, $\kappa^{t}$, $\Phi^{t}$ and $\Omega^{t}$ are respectively the surface gravity, electric potential and   angular velocity of  the horizon; $Q_{h}~,~A_{h}$ and $J_{h}$ are respectively the charge, area and angular momentum of  the horizon. The label '$t$' denotes the particular time evolution field ($t^{\mu}$)  associated with the spatial hypersurface chosen.  $E^{t}_{h}$ is  assumed here to be a function of $A_{h}$, $Q_{h}$ and $J_{h}$.

 As argued in \cite{aspm 16} , mass can be defined on the isolated horizon.

\subsection{Quantum Black Hole Algebra And Quantum Geometry}

Like for all quantum systems, an operator algebra of fundamental observables is required to have a proper quantum description of black holes. Classically, generic black holes are represented by four parameters $ (M,Q,J,A) $, with three of them independent. It is not possible to have a black hole with $ M = 0$ and $ Q,J \neq 0 $. So, additional structures i.e. charge and angular momentum are fundamental observables in a quantum theory. We choose area $(A)$ as the third fundamental observable. So, Mass($M$) becomes the secondary observable i.e. $M=M(A,Q,J)$. So, the algebraic approach of black hole quantization gives, $\widehat {Q} ,\widehat {J}, \widehat {A} $ as quantum operators of fundamental observables and $\widehat{M}(\widehat{H}_{b})$ as quantum operator of seceondary observable. All these correspond to the isolated horizon of a black hole.\\

 In Loop Quantum Gravity(LQG), quantum black holes are represented by spin network, collection of graphs with links and vertices \cite{rosm}. Spin networks are duals of cellular decompositions of space, where a certain volume is associated to a vertex and each boundary area with certain links. So, geometry of a black hole horizon is completely determined by the intersections of the graphs with its boundary. These intersections are labelled  with $p \in  N^{+} = (1, 2, 3, ...)$ and each link is assigned to go through $p$ th color $j_{p} \in  N/2 = (0, 1/2, 1, 3/2, ...)$, quantum geometry of the surface  is characterized by a p-tuple of spins $j = (j_{1} , ..., j_{p} )$ . A system of $n$ particles each having a spin $j_{p}$ with states in a single-particle tensor product Hilbert space
$ \mathcal{H}_{b} = \mathcal{H}_{b}^{(j_{1})}  \otimes ... \otimes \mathcal{H}_{b}^{(j_{n})}$ . Simultaneous eigenstates of the $i$th component $\widehat{J}^{i}_{p}$ of the angular momentum operator$(\widehat{J}_{p}) $ and of the Casimir operator $(\widehat{J}^{2}_{p})$ constitute an orthonormal basis for $\mathcal{H}_{b}$. These states are the spin network states. $\widehat{J}^{i}_{p}$ and $\widehat{J}^{2}_{p}$ have eigenvalues $m_{p}$ and $j_{p}(j_{p}+1)$ respectively, where $m_{p}$ is the spin projection quantum number of the $p$th link, can take on the values $ (-j_{p}, -j_{p} + 1 , ..., j_{p}-1, j_{ p} )$. So, spin network states can be explicitly denoted  as $\vert (j_{p}, m_{p})^{n}_{1},...\rangle$ , with $n=p_{max}$.\\

Now, LQG gives the action of black hole horizon area operator $(\widehat{A})$ and angular momentum operator $(\widehat{J})$ respectively as \cite{rdcr},
\begin{equation}
\widehat{A}\vert (j_{p}, m_{p})^{n}_{1},...\rangle = A \vert (j_{p}, m_{p})^{n}_{1},...\rangle = 8\pi l_{p}^{2}\gamma \sum_{p=1}^{n} m_{p} \sqrt{j_{p}(j_{p}+1)}\vert (j_{p}, m_{p})^{n}_{1},...\rangle \label{areope}
\end{equation}
Where, $A=8\pi l_{p}^{2}\gamma \sum_{p=1}^{n} m_{p} \sqrt{j_{p}(j_{p}+1)}\equiv $ area of black hole horizon  , $l_{p} \equiv$ planck length and $\gamma \equiv $Immirzi parameter.
\begin{equation}
\sum_{p=1}^{n} m_{p}\widehat{J}^{i}_{p}\vert (j_{p}, m_{p})^{n}_{1},...\rangle = \frac{J}{l_{p}^{2}\gamma}\delta^{i}_{1}\vert (j_{p}, m_{p})^{n}_{1},...\rangle = \delta^{i}_{1}\sum_{p=1}^{n} m_{p}\vert (j_{p}, m_{p})^{n}_{1},...\rangle \label{anumomope}
\end{equation}
Where, $J= l_{p}^{2}\gamma \sum_{p=1}^{n} m_{p}\equiv $ Angular momentum of the black hole.\\

It is physically obvious that both area and charge should be invariant under SO(3) rotations and that the area should also be U(1) gauge invariant. Since the angular momentum is a measure for rotation($SO(3)$ Group) and the charge is the generator of the U(1) global gauge group . These give,
\begin{equation}
[\widehat{A},\widehat{J}]= [\widehat{A},\widehat{Q}]= [\widehat{Q},\widehat{J}]=0 \label{commu1}
\end{equation}
Since $\widehat{M}(\widehat{H}_{b})$ is a quantum operator of secondary observable ($M(A,J,Q)$), Equ no. \ref{commu1} can be extended as,
\begin{equation}
[\widehat{A},\widehat{J}]= [\widehat{A},\widehat{Q}]= [\widehat{A},\widehat{M}]= [\widehat{Q},\widehat{J}]= [\widehat{M},\widehat{Q}]=[\widehat{J},\widehat{M}]=0 \label{commu2}
\end{equation}
The generic quantum black hole horizon(boundary) state is denoted as, $\vert j_{p}, m_{p}, q \rangle$, where $eq$ is the eigenvalue of the charge operator($\widehat{Q}$) with $q$ is a integer no. and $e$ is the fundamental $U(1)$ charge.\\

Equ no. \ref{commu2} implies that $\vert j_{p}, m_{p}, q \rangle$ is a simultaneous eigenstate of $\widehat{A},\widehat{J},\widehat{Q},\widehat{M}$ with eigenvalues as follows,\\

 \hspace{1.2 in} $\widehat{A}\vert j_{p}, m_{p}, q \rangle=A\vert j_{p}, m_{p}, q \rangle,$ \hspace{.2 in} $l_{p}\sqrt{\gamma}\widehat{Q}\vert j_{p}, m_{p}, q \rangle=Q\vert j_{p}, m_{p}, q \rangle$  
\begin{equation}
l_{p}^{2}\gamma\vert j_{p}, m_{p}, q \rangle=J\vert j_{p}, m_{p}, q \rangle,  \hspace{.2 in} l_{p}\sqrt{\gamma}\widehat{M}\vert j_{p}, m_{p}, q \rangle=M\vert j_{p}, m_{p}, q \rangle
\end{equation}
Where, $Q=l_{p}\sqrt{\gamma}eq \equiv$ charge of the black hole, $M \equiv$ mass of the black hole and rest are as before. \\

The Hilbert space of  a generic quantum spacetime is given as, $\mathcal{H}=\mathcal{H}_{b}{\otimes}\mathcal{H}_{v}$ , where $b(v)$ denotes the boundary (bulk)  space. A generic quantum state is  thus given as
\begin{equation}
\vert\Psi\rangle=\sum\limits_{b,v} C_{b,v} \vert\chi_{b}\rangle {\otimes} \vert\psi_{v}\rangle ~\label{genstate} 
\end{equation} 
Now, the full Hamiltonian operator ($\widehat{H}$),  operating on $\mathcal{H}$ is given by
\begin{equation}\label{hamil}
\widehat{H}\vert\Psi\rangle=(\widehat{H_{b}}{\otimes}I_{v}+I_{b}{\otimes}\widehat{H_{v}})\vert\Psi\rangle
\end{equation} 
where,  respectively, $I_{b} (I_{v})$ are identity operators on $\mathcal{H}_{b} (\mathcal{H}_{v})$ and $\widehat{H_{b}} (\widehat{H_{v}})$ are the Hamiltonian operators on $\mathcal{H}_{b}(\mathcal{H}_{v})$. \\

The Hilbert space of  a generic quantum spacetime is given as, $\mathcal{H}=\mathcal{H}_{b}{\otimes}\mathcal{H}_{v}$ , where $b(v)$ denotes the boundary (bulk)  space. A generic quantum state is  thus given as
\begin{equation}
\vert\Psi\rangle=\sum\limits_{b,v} C_{b,v} \vert\chi_{b}\rangle {\otimes} \vert\psi_{v}\rangle ~\label{genstate} 
\end{equation} 
Now, the full Hamiltonian operator ($\widehat{H}$),  operating on $\mathcal{H}$ is given by
\begin{equation}\label{hamil}
\widehat{H}\vert\Psi\rangle=(\widehat{H_{b}}{\otimes}I_{v}+I_{b}{\otimes}\widehat{H_{v}})\vert\Psi\rangle
\end{equation} 
where,  respectively, $I_{b} (I_{v})$ are identity operators on $\mathcal{H}_{b} (\mathcal{H}_{v})$ and $\widehat{H_{b}} (\widehat{H_{v}})$ are the Hamiltonian operators on $\mathcal{H}_{b}(\mathcal{H}_{v})$. \\

The first class constraints are realized on Hilbert space as annihilation constraints on physical states. The bulk Hamiltonian operator thus annihilates bulk physical states
\begin{eqnarray}
\widehat{H_{v}}\vert\psi_{v}\rangle=0 \label{bulkham}
\end{eqnarray}\\

Any generic quantum bulk Hilbert space is invariant under local $U(1)$ gauge transformations and local spacetime rotations (the latter, as part of local Lorentz invariance). Since $ \widehat{Q_{v}} , \widehat{J_{v}}$  are the generators of $U(1)$ gauge transformation and local spacetime rotation for bulk spacetime respectively, they individually annihilates bulk states i.e. $ \widehat{Q_{v}} | \psi_{v} \rangle = 0 ,$ \hspace{.03 in} $ \widehat{J_{v}}] | \psi_{v} \rangle = 0$.\\

So, for generic bulk states
\begin{eqnarray}
[\widehat{H_{v}} - \Phi \widehat{Q_{v}} - \Omega \widehat{J_{v}}] | \psi_{v} \rangle = 0 ~.~\label{fullham}
\end{eqnarray}

\subsection{Grand Canonical Partition Function}

Consider the black hole immersed in a heat bath, at some (inverse) temperature $\beta$, with which it can exchange energy, charge and angular momentum. The grand canonical partition function of the black hole is given as,

\begin{eqnarray}
Z_{G}=Tr(exp(-\beta\widehat{H}+\beta\Phi\widehat{Q}+\beta\Omega\widehat{J})) ~\label{gcpf}
\end{eqnarray}
where  the trace is taken over all states. This definition,  together with eqn.s (\ref{genstate}) and (\ref{fullham}), yields
\begin{eqnarray}
Z_{G} &=& \sum_{b,v} \vert C_{b,v} \vert^{2} \langle\psi_{v}\vert\psi_{v}\rangle \langle\chi_{b}\vert  exp(-\beta\widehat{H}+\beta\Phi\widehat{Q}+\beta\Omega\widehat{J})  \vert\chi_{b}\rangle \nonumber \\ 
&=& \sum\limits_{b} \vert C_{b} \vert^{2} \langle\chi_{b}\vert exp(-\beta\widehat{H}+\beta\Phi\widehat{Q}+\beta\Omega\widehat{J})  \vert\chi_{b}\rangle ~,~ \label{pf+ham}
\end{eqnarray}
assuming that the bulk states are normalized. The partition function  thus turns out to be  completely determined by the boundary states ($Z_{Gb}$), i.e.,
\begin{eqnarray}
Z=Z_{Gb} &=& Tr_{b} \exp(-\beta\widehat{H}+\beta\Phi\widehat{Q}+\beta\Omega\widehat{J}) \nonumber \\
&=& \sum\limits_{k,l,m} g(k,l,m) \hspace{.1 in} \exp(-\beta(E(A_{k},Q_{l},J_{m})-\Phi Q_{l}-\Omega J_{m})) ~,~\label{bdypf}
\end{eqnarray}
where $g(k,l,m)$ is the degeneracy corresponding to energy $E(A_{k},Q_{l},J_{m})$ and $k,l,m$ are the quantum numbers corresponding to  eigenvalues of area, charge and angular momentum respectively. The application of the Poisson resummation formula \cite{cm04} gives
\begin{equation}
Z_{G}=\int dx\hspace{.05 in} dy\hspace{.05 in} dz\hspace{.05 in}g(A(x),Q(y),J(z))\hspace{.05 in}  \exp(-\beta(E(A(x),Q(y),J(z))-\Phi Q(y)-\Omega J(z)))
\end{equation}
where $x,y,z$ are respectively the continuum limit of $k,l,m$ respectively. 

A change of variables gives, 
\begin{equation}
Z_{G}=\int dA\hspace{.05 in} dQ\hspace{.05 in} dJ\hspace{.05 in} \exp [S(A)-\beta(E(A,Q,J)-\Phi Q-\Omega J)]~,~ \label{pfresult}
\end{equation}
where, following \cite{ll}, the {\it microcanonical} entropy $(S(A))$ of the horizon is defined by  \hspace{1 in} $\exp S(A) \equiv \frac{ g(A(x),Q(y),J(z))}{\frac{dA}{dx}\frac{dQ}{dy}\frac{dJ}{dz}}$.

\subsection{Saddle Point Approximation and Stability Criteria}

 The equilibrium configuration of black hole is given by the saddle point $(\bar{A},\bar{Q},\bar{J})$ in the three dimensional space of integration over area, charge and angular momentum with fluctuations $a=(A-\bar{A}), q=(Q-\bar{Q}),j=(J-\bar{J})$ around the saddle point.  Taylor expanding eqn (\ref{pfresult}) about the saddle point, yields 
\begin{eqnarray}
Z_{G} &=& \exp[ S(\bar{A})-\beta M(\bar{A},\bar{Q},\bar{J})+\beta\Phi \bar{Q}+\beta\Omega\bar{J}] \nonumber \\
&\times & \int da~ dq~ dj~ \exp \{-\frac{\beta}{2}[( M_{AA}-\frac{S_{AA}}{\beta} )a^{2} + ( M_{QQ})q^{2}+(2 M_{AQ})aq \nonumber \\
&+& ( M_{JJ})j^{2}+(2M_{AJ})aj+(2 M_{QJ})qj] \} ~ \label{sadpt}
\end{eqnarray}
where  $M_{JJ} = \frac{\partial^{2}M}{\partial J^{2}} \big\vert_{(\bar{A},\bar{Q},\bar{J})}$  etc. as described in \cite{aspm 16}. \\

Convergence of the integral (\ref{sadpt}) implies  that the Hessian matrix ($H$) has to be positive definite, where
\begin{eqnarray}
 H = \left( \begin{array}{ccc}
\beta M_{AA}(\bar{A},\bar{Q},\bar{J})- S_{AA}(\bar{A}) \hspace{.3 in}& \beta M_{AQ}(\bar{A},\bar{Q},\bar{J}) \hspace{.3 in} & \beta M_{AJ}(\bar{A},\bar{Q},\bar{J})\vspace{.1 in} \\  
\beta M_{AQ}(\bar{A},\bar{Q},\bar{J}) \hspace{.3 in}& \beta M_{QQ}(\bar{A},\bar{Q},\bar{J}) \hspace{.3 in}& \beta M_{JQ}(\bar{A},\bar{Q},\bar{J}) \vspace{.1 in} \\
\beta M_{AJ}(\bar{A},\bar{Q},\bar{J}) \hspace{.3 in}& \beta M_{JQ}(\bar{A},\bar{Q},\bar{J}) \hspace{.3 in}& \beta M_{JJ}(\bar{A},\bar{Q},\bar{J}) \end{array} \right) \label{hess}
\end{eqnarray}
The necessary and sufficient conditions for  a real symmetric square matrix to be positive definite is : 'determinants all principal square submatrices, and the determinant of the full matrix, are positive.'\cite{meyer} This condition leads to the  `stability criteria'  that are described in \cite{aspm 16}.Ofcourse, (inverse) temperature $\beta$ is assumed to be positive for a stable configuration.\\  

 Since, we are considering quantum theory of gravity, we have to consider the effect of quantum spacetime fluctuations on microcanonical entropy of isolated horizons . It has been shown that \cite{km00} the microcanonical entropy for  {\it macroscopic} isolated horizons($S$) has the form

\begin{eqnarray}
S~&=&~S_{BH} ~-~\frac32 \log S_{BH} +{\cal O}(S_{BH}^{-1}) ~\label{kment} \\
S_{BH} ~&=& ~ \frac{A}{4 A_P}~,~A_P \rightarrow {\rm Planck~area} ~. \label{bek}
\end{eqnarray} \\

In reference \cite{km00} , the formula \ref{kment} was derived for non-rotating black holes in four dimensional spacetime.This is based on a three dimensional SU(2) Chern-Simons theory. Where as consideration of U(1) theory gives, $-\log (S_{BH})$. Although the cofficeients mismatch with each other, but both are logarithmic corrections and subdominating for large horizon area. The detail study of isolated horizon for rotating black holes \cite{acj} shows that many properties of rotating black hole are like that of non-rotating ones. This hints towards the possibility of similar \ref{kment}  correction for microcanonical entropy of rotating black holes as well. The approach, of viewing Hawking radiation from a black hole as quantum tunneling of particles through the event horizon, shows \cite{aksa} that microcanonical entropy of Kerr-Newman black hole has form similar to that of \ref{kment} .  In the reference \cite{asen}, it is extensively shown that for various types of black holes in various spacetime dimension with various charges and angular momentums, the corrections of black hole entropy are mostly lagarithmic and hence subdominating for large black hole area. So, the exact form for correction of microcanonical entropy really does not alter any calculation of the rest of the paper as we will only bother about the leading order values in large area limit($A>>A_{P}$).

\section{Thermal Fluctuation And correlation Among Hairs Of ADS Kerr-Newman Black Hole}

 The expectation value of fluctuation of any quantitity is the standard deviation of that quantity. It is a statistical measure of deviation for any distribution. The knowledge of probability theory and the last expression of grand canonical partition function (\ref{sadpt}) together give the standard deviation of charge($Q$) as,
 \begin{eqnarray}
(\Delta Q)^{2} = \frac{\int da~ dq~ dj~  q^{2}\exp \{-\frac{\beta}{2}[( M_{AA}-\frac{S_{AA}}{\beta} )a^{2} + ( M_{QQ})q^{2}+(2 M_{AQ})aq +( M_{JJ})j^{2}+(2M_{AJ})aj+(2 M_{QJ})qj] \}}{\int da~ dq~ dj~   \exp \{-\frac{\beta}{2}[( M_{AA}-\frac{S_{AA}}{\beta} )a^{2} + ( M_{QQ})q^{2}+(2 M_{AQ})aq + ( M_{JJ})j^{2}+(2M_{AJ})aj+(2 M_{QJ})qj] \}} ~ ~ ~ \label{chflu}
\end{eqnarray}
where, $\Delta Q$ is the standard deviation of black hole charge. Similarly, $\Delta A$ and $\Delta J$ are defined for horizon area and angular mementum of the black hole.\\

The correlation function between charge($Q$) and angular momentum ($J$) is denoted as $\Delta QJ$ and is defined as, 
\begin{eqnarray}
\Delta QJ = \frac{\int da~ dq~ dj~  qj\exp \{-\frac{\beta}{2}[( M_{AA}-\frac{S_{AA}}{\beta} )a^{2} + ( M_{QQ})q^{2}+(2 M_{AQ})aq +( M_{JJ})j^{2}+(2M_{AJ})aj+(2 M_{QJ})qj] \}}{\int da~ dq~ dj~   \exp \{-\frac{\beta}{2}[( M_{AA}-\frac{S_{AA}}{\beta} )a^{2} + ( M_{QQ})q^{2}+(2 M_{AQ})aq + ( M_{JJ})j^{2}+(2M_{AJ})aj+(2 M_{QJ})qj] \}} ~ ~ ~ \label{camcor}
\end{eqnarray}
Similarly, $\Delta QA$ and $\Delta JA$ are defined for the black hole.\\

The expression(\ref{sadpt}) and (\ref{chflu}) together give,
\begin{eqnarray}
(\Delta Q)^{2} &=& -\frac{2}{\beta}\cdot\frac{1}{ Z_{G}}\cdot\frac{\partial Z_{G}}{\partial M_{QQ}}  \label{chflu1}
\end{eqnarray}
Similarly, $(\Delta A)^{2}$ , $(\Delta J)^{2}$, $\Delta QA$ , $\Delta JA$ , $\Delta QJ$ are defined by taking partial derivatives with respect to $ (M_{AA}-\frac{S_{AA}}{\beta}) $ , $M_{JJ}$, $M_{QA}$, $M_{JA}$ and $M_{QJ}$ respectively i.e.
\begin{eqnarray}
(\Delta A)^{2} &=& -\frac{2}{\beta}\cdot\frac{1}{ Z_{G}}\cdot\frac{\partial Z_{G}}{\partial (M_{AA}- \frac{S_{AA}}{\beta})}  \label{arflu}
\end{eqnarray}
\begin{eqnarray}
(\Delta J)^{2} &=& -\frac{2}{\beta}\cdot\frac{1}{ Z_{G}}\cdot\frac{\partial Z_{G}}{\partial M_{JJ}}  \label{amflu}
\end{eqnarray}
\begin{eqnarray}
\Delta QA &=& -\frac{1}{\beta}\cdot\frac{1}{ Z_{G}}\cdot\frac{\partial Z_{G}}{\partial M_{QA}}  \label{cacor}
\end{eqnarray}
\begin{eqnarray}
\Delta JA &=& -\frac{1}{\beta}\cdot\frac{1}{ Z_{G}}\cdot\frac{\partial Z_{G}}{\partial M_{JA}} \label{amacor}
\end{eqnarray}
\begin{eqnarray}
\Delta QJ &=& -\frac{1}{\beta}\cdot\frac{1}{ Z_{G}}\cdot\frac{\partial Z_{G}}{\partial M_{QJ}}  \label{camcor1}
\end{eqnarray}
Equation no. (\ref{sadpt}), (\ref{hess}) and (\ref{chflu1}) together give,
\begin{eqnarray}
(\Delta Q)^{2} = \frac{2}{\vert H \vert}\cdot \Big((\beta M_{AA}- S_{AA})\cdot \beta M_{JJ} - (\beta M_{AJ})^{2}\Big)\label{chflu2}
\end{eqnarray}
where, $\vert H \vert$ is the determinant of the hessian matrix($H$). \\

Equation no. (\ref{sadpt}), (\ref{hess}) and (\ref{arflu}) together give,
\begin{eqnarray}
(\Delta A)^{2} = \frac{2}{\vert H \vert}\cdot \Big(\beta ^{2} \big(M_{QQ}M_{JJ} - (M_{JQ})^{2}\big)\Big) \label{arflu1}
\end{eqnarray}
Equation no. (\ref{sadpt}), (\ref{hess}) and (\ref{amflu}) together give,
\begin{eqnarray}
(\Delta J)^{2} = \frac{2}{\vert H \vert}\cdot \Big((\beta M_{AA}- S_{AA})\cdot \beta M_{QQ} - (\beta M_{AQ})^{2}\Big) \label{amflu1}
\end{eqnarray}
Equation no. (\ref{sadpt}), (\ref{hess}) and (\ref{cacor}) together give,
\begin{eqnarray}
\Delta QA = \frac{2}{\vert H \vert}\cdot \Big(\beta^{2}(M_{AQ}M_{JJ} - M_{JQ}M_{AJ})\Big) \label{cacor1}
\end{eqnarray}
Equation no. (\ref{sadpt}), (\ref{hess}) and (\ref{amacor}) together give,
\begin{eqnarray}
\Delta JA =  \frac{2}{\vert H \vert}\cdot \Big(\beta^{2}(M_{AJ}M_{QQ} - M_{JQ}M_{AQ})\Big) \label{amacor1}
\end{eqnarray}
Equation no. (\ref{sadpt}), (\ref{hess}) and (\ref{camcor1}) together give,
\begin{eqnarray}
\Delta QJ = \frac{2}{\vert H \vert}\cdot \Big(\beta M_{JQ}(\beta M_{AA} - S_{AA}) - \beta^{2} M_{AQ}M_{AJ}\Big) \label{camcor2}
\end{eqnarray}

The AdS Kerr-Newman black hole is given in Boyer–Lindquist coordinates as 
\begin{equation}\label{adsknmetric}
ds^{2}= -\frac{\Delta_{r}}{\rho^{2}}(dt-\frac{a\hspace{.02 in} sin^{2}\theta}{\Sigma}\hspace{.02 in} d\phi)^{2} +\frac{\Delta_{\theta}\hspace{.02 in} sin^{2}\theta}{\rho^{2}}(\frac{r^{2}+a^{2}}{\Sigma} d\phi -a\vspace{.02 in}dt)^{2} +\frac{\rho^{2}}{\Delta_{r}}dr^{2}+ \frac{\rho^{2}}{\Delta_{\theta}} d\theta ^{2}
\end{equation}
where, $ \Sigma= 1-\frac{a^{2}}{l^{2}},\hspace{.1 in} \Delta_{r}=(r^{2}+ a^{2})(1+\frac{r^{2}}{l^{2}})-2\hspace{.02 in}M \hspace{.02 in}r +Q^{2} ,\hspace{.1 in}\Delta_{\theta}= 1-\frac{a^{2}cos^{2}\theta}{l^{2}},\hspace{.1in} \rho^{2}= r^{2}+a^{2}\hspace{.02 in}cos^{2}\theta  , \hspace{.1 in}  a=\frac{J}{M} $. The generalized Smarr formula for the AdS Kerr-Newman Black Hole is given as \cite{cck} 
\begin{equation}\label{adsknmass}
M^{2}=\frac{A}{16\pi}+\frac{\pi}{A}(4J^{2}+Q^{4})+\frac{Q^{2}}{2}+\frac{J^{2}}{l^{2}}+\frac{A}{8\pi l^{2}}(Q^{2}+\frac{A}{4\pi}+\frac{A^{2}}{32\pi^{2}l^{2}})
\end{equation}
where  the cosmological constant ($\Lambda$) is defined in terms of a cosmic length parameter as $\Lambda = -1/l^2$.

As before, our interest is in astrophysical (macroscopic) charged, rotating black holes whose horizon area exceeds by far the Planck area. It has shown \cite{aks} that, ADS kerr-newman black holes are stable if $\frac{A}{l^{2}}>> \frac{Q^{2}}{A}, \frac{J}{A}$. So, we can approximate (\ref{adsknmass}) as follows 
\begin{equation}\label{adsknmassapp}
M\thickapprox\frac{A^{3/2}}{16\pi^{3/2}l^{2}}+\frac{A^{1/2}}{4\pi^{1/2}}+\frac{\pi^{1/2}Q^{2}}{A^{1/2}}+\frac{8\pi^{3/2}J^{2}}{A^{3/2}} ~.
\end{equation}
 Equation no. (\ref{hess}), (\ref{chflu2}) and (\ref{adsknmassapp}) together give,
 \begin{eqnarray}
(\Delta Q)^{2} \approx \frac{3 A_{p}A}{8 \pi^{2}l^{2}} \label{chflu3}
\end{eqnarray}
Equation no. (\ref{hess}), (\ref{arflu1}) and (\ref{adsknmassapp}) together give,
 \begin{eqnarray}
(\Delta A)^{2} \approx 16A_{p}A \label{arflu2}
\end{eqnarray}
Equation no. (\ref{hess}), (\ref{amflu1}) and (\ref{adsknmassapp}) together give,
 \begin{eqnarray}
(\Delta J)^{2} \approx \frac{3 A_{p}A^{2}}{64 \pi^{3}l^{2}} \label{amflu2}
\end{eqnarray}
Equation no. (\ref{hess}), (\ref{cacor1}) and (\ref{adsknmassapp}) together give,
\begin{eqnarray}
\Delta QA \approx -8A_{P}Q \label{cacor2}
\end{eqnarray}
Equation no. (\ref{hess}), (\ref{amacor1}) and (\ref{adsknmassapp}) together give,
\begin{eqnarray}
\Delta JA \approx -24A_{P}J \label{amacor2}
\end{eqnarray}
Equation no. (\ref{hess}), (\ref{camcor2}) and (\ref{adsknmassapp}) together give,
\begin{eqnarray}
\Delta QJ \approx  \frac{-12A_{P}JQ}{A} \label{camcor3}
\end{eqnarray}
 Ofcourse, last six expressions are only the leading order terms in large horizon area limit.\\

 It is extremely interesting to note that $((\Delta J)^{2})$ and $((\Delta Q)^{2})$ are independent of $J$ and $Q$ respectively. It means for a large black hole there are finite amount of fluctuations of charge and angular momentum even for a almost neutral, nonrotating ADS black hole. \\
 
The measure of fluctuations of the above six fluctuations are given as,\\
 
$1.$ Measure of Area fluctuation
\begin{eqnarray}
 \frac{\Delta A}{A} \approx  4 \sqrt{\frac{A_{P}}{A}} \label{arflu3}
\end{eqnarray}\ 

$2.$ Measure of Charge fluctuation
\begin{eqnarray}
 \frac{\Delta Q}{Q} \approx  \sqrt{\frac{3}{8\pi^{2}}} \cdot {\frac{\sqrt {A_{P}A}}{Ql}} \label{chflu3}
\end{eqnarray}\
 
 $3.$ Measure of Angular Momentum fluctuation
\begin{eqnarray}
 \frac{\Delta J}{J} \approx  \sqrt{\frac{3}{64\pi^{3}}} \cdot \sqrt{\frac{A_{P}A^{2}}{J^{2}l^{2}}} \label{amflu3}
\end{eqnarray}\

$4.$ Measure of Charge - Area correlation
\begin{eqnarray}
 \sqrt{\frac{|\Delta QA|}{QA}} \approx  \sqrt{\frac{8A_{P}}{A}} \label{cacor3}
\end{eqnarray}\

$5.$ Measure of Charge - Angular Momentum correlation
\begin{eqnarray}
 \sqrt{\frac{|\Delta QJ|}{QJ}} \approx  \sqrt{\frac{12A_{P}}{A}} \label{camcor4}
\end{eqnarray}\

$6.$ Measure of Area - Angular Momentum correlation
\begin{eqnarray}
 \sqrt{\frac{|\Delta AJ|}{AJ}} \approx  \sqrt{\frac{24A_{P}}{A}} \label{aamcor3}
\end{eqnarray}\

Equation Nos. (\ref{arflu3}), (\ref{cacor3}) , (\ref{camcor4}) , (\ref{aamcor3}) imply respectively that Measure of Area fluctuation ($\frac{\Delta A}{A}$) ,  Charge - Area correlation \Big($\sqrt{\frac{|\Delta QA|}{QA}}$ \Big) ,  Charge - Angular Momentum correlation \Big($\sqrt{\frac{|\Delta QJ|}{QJ}}$\Big) and   Area - Angular Momentum correlation \Big($\sqrt{\frac{|\Delta AJ|}{AJ}}$\Big) are vanishly small for large black holes ($A>>A_{P}$). These results imply that we are dealing not only around equilibrium point but also it is a stable equilibrium point. The surprising fact is that these measures of correlations are independent of charge ($Q$) , angular momentum ($J$) of the black hole. \\

The expression no.(\ref{adsknmassapp}) implies that for large black hole area($A$) , \
\begin{eqnarray}
\frac{A^{3/2}}{16\pi^{3/2}l^{2}} > \frac{\pi^{1/2}Q^{2}}{A^{1/2}}  ,\hspace{.2 in}  \frac{A^{3/2}}{16\pi^{3/2}l^{2}} > \frac{8\pi^{3/2}J^{2}}{A^{3/2}} \label{compare}
\end{eqnarray}\
Equation nos. (\ref{chflu3})(\ref{amflu3}) and (\ref{compare}) together give,\
\begin{eqnarray}
 \frac{\Delta Q}{Q} >  \sqrt\frac{6A_{P}}{A} \label{chflu4}
\end{eqnarray}\
\begin{eqnarray}
 \frac{\Delta J}{J} >  \frac{1}{64\pi^{3}} \sqrt\frac{3A_{P}}{2A} \label{amflu4}
\end{eqnarray}\
Last two expressions imply that measure of charge and angular momentum fluctuations are vanishly small for large black holes($A>>A_{P}$). Although last two expressions of (\ref{chflu4}) and (\ref{amflu4}) are the lower bounds, but still they are independent of charge($Q$) and angular momentum($J$) of the black hole and eventually are zero in large black hole limit.

\section{Summary and Discussion}
We reiterate that our analysis is quite independent of specific classical spacetime geometries, relying as it does on quantum aspects of spacetime. The construction of the partition function used standard formulations of equilibrium statistical mechanics augmented by results from canonical Quantum Gravity, with extra inputs regarding the behaviour of the microcanonical entropy as a function of area {\it beyond the Bekenstein-Hawking area law}, as for instance derived from Loop Quantum Gravity \cite{km00}. We use classical metric only as an input which gives the dependence of mass of black hole($M$) on its' charge($Q$), area($A$) and angular momentum($J$).\\

In large horizon area limit, it turns out that  for a quantum ADS black hole , leading order fluctuations of charge($(\Delta Q)^{2}$) and angular momentum($(\Delta J)^{2}$) are independent of its' charge($Q$) and angular momentum($J$). This implies even a black hole with vanishingly small charge($Q$) and angular momentum($J$) can have finite fluctuations in respective quantities. Our analysis can be trivially extended for black holes with any number of hairs in any space time dimension \cite{aks}.  The $S_{AA}$ term is present everywhere in the calculation. The non vanishing contribution of this term is pure artifect of quantum fluctuation of spacetime. Thermal fluctuations are present along with this as we are considering black hole to be immersed in a extended thermal bath. So, thermal fluctuations and correlations , that we have calculated , take care of quantum fluctuation of spacetime within it automatically. This is extremely interesting in its' own merit. We choose this example of quantum ADS black hole as  AdS/CFT correspondence tells that string theory on AdS space is dual to a conformal field theory (CFT) on the boundary of that AdS space \cite{sc} \cite{jmm}. It has also been shown using the AdS/CFT correspondence that the asymptotically AdS black hole is dual to a strongly coupled gauge theory at finite temperature \cite{jd} - \cite{am}. It is possible to study the strongly correlated condensed-matter physics using the AdS/CFT correspondence.  Holographic model of superconductors has also been constructed from blackhole solutions using the AdS/CFT correspondence \cite{apt}. Hence our results on quantum ADS black hole may have some imprints on  possible applications for the strongly correlated condensed-matters systems.

\end{document}